# Strength In Diversity: Small Bodies as the Most Important Objects in Planetary Sciences

A White Paper submitted to the National Academy of Sciences Planetary and Astrobiology Decadal Survey 2023-2032

Laura M. Woodney (California State University, San Bernardino)
Andrew S. Rivkin (Johns Hopkins University Applied Physics Laboratory)
Walter Harris (University of Arizona, Tucson)
Barbara A. Cohen (NASA GSFC)
Gal Sarid (SETI Institute)
Maria Womack (Florida Space Institute, University of Central Florida)
Olivier Barnouin (Johns Hopkins University Applied Physics Laboratory)
Kat Volk (University of Arizona, Tucson)
Rachel Klima (Johns Hopkins University Applied Physics Laboratory)
Yanga R. Fernandez (University of Central Florida)
Jordan K. Steckloff (Planetary Science Institute, University of Texas at Austin)
Paul A. Abell (NASA Johnson)

**Executive Summary**
The small bodies of the solar system are the unaccreted leftovers of planetary formation. In planetary science, they could also be mistaken for leftovers in the sense that they are "everything else" after the planets and their satellites (or sometimes just their "regular" satellites) are accounted for. This mistaken view elides the great diversity of compositions, histories, and present-day conditions and processes found in the small bodies, and the interdisciplinary nature of their study. Understanding small bodies is critical to planetary science as a field, and we urge planetary scientists and our decision makers to continue to support science-based mission selections and to recognize that while small bodies have been grouped together for convenience, *the diversity of these objects in terms of composition, mass, differentiation, evolution, activity, dynamical state, physical structure, thermal environment & history, and formation* ***vastly exceeds*** *the observed variability in the major planets and their satellites.* Treating them as a monolithic group with interchangeable members does a grave injustice to the range of fundamental questions they address. We advocate for a deep and ongoing program of missions, telescopic observations, R&A funding, and student support that respects this diversity.

**What are the small bodies?**
In practice, the small bodies comprise every physical object in the solar system that isn't a planet or a moon. This includes most obviously the asteroids (main belt, near-Earth) and comets (Jupiter Family, Halley-Type, Oort Cloud), but also includes Centaurs, trans-Neptunian objects, objects still in the Oort Cloud, irregular satellites (including Deimos and Phobos), planetary trojans, and interstellar objects. The largest of the small bodies are the dwarf planets: Ceres, Eris, Pluto, and a large number of additional objects in the outer solar system that approach the (sometimes controversial) line separating planets from non-planets. Interplanetary dust is typically derived from comets and asteroids and its study, along with the study of meteorites and their delivery to Earth, is generally done in the context of small bodies research.

The differences between small body populations are not merely dynamical but also in their records of presolar processes, compositions, and mechanisms of planetesimal formation. Their surfaces and interior structures range from near primordial to highly processed, leading to as wide an array of evidence of physical evolution across their size and dynamical range as that which differentiates these processes on Mercury, Venus, Earth, and Mars. Consider that the range of properties covered by small bodies is astronomical, even for literally astronomical objects. Eris and Pluto have masses $\sim 10^{22}$ kg and Quaoar and Ceres $\sim 10^{21}$ kg, while Itokawa and Bennu are $\sim 10^{10}$ kg. The smallest numbered asteroid, 478784 (2012 UV136), likely has a mass of $\sim 10^7$ kg, roughly the same size and mass as the Chelyabinsk impactor and of the smallest sungrazing comets seen by SOHO. The 15 orders of magnitude in mass represented by this range, even ignoring smaller meteoroids (let alone dust), is larger than any group of objects called by the same name, and is larger than the 12 orders of magnitude in mass separating Neptune from the supermassive black hole at the center of the Milky Way galaxy. With this perspective, it is hardly

| Table1: Diversity of Small Solar System Bodies ||
|---|---|
| **Object Type**: *Properties* | **Primary Science Objectives** |
| **Trans Neptunian Objects (TNO)**: *Minimally processed, undifferentiated, volatile rich planetesimal.* | Cold circumstellar disk accretion processes, composition, temperature, and timing. Binary system formation. Planetary Migration. |
| **Trans Neptunian Dwarf Planets**: *Hydrostatic volatile dominated protoplanet.* | Planetary formation. Differentiation processes. Ocean world potential. Cryogenic tectonics & volcanism. Sublimation based atmospheres. |
| **Centaurs**: *Objects in dynamical transition between the inner and outer solar system.* | Early stage physical evolution. Non-water based volatile activity. Intermediate state for physical processes between TNOs & JFCs. |
| **Jupiter Family Comets (JFC)**: *Highly processed undifferentiated planetesimal* | Link back to circumstellar disk abundances and planetesimal accretion processes. Volatile retention and loss mechanisms, resurfacing processes. |
| **Long Period Comets (LPC) - Halley-Type and Oort Cloud**: *Partially processed undifferentiated planetesimal* | Link back to accretion in and composition of warm circumstellar disk. Volatile retention and loss mechanisms, resurfacing processes. Transition between water and non-water based activity. |
| **Main Belt Comets (MBC)**: *End-state mostly inactive volatile rich planetesimal.* | Surface processing, geologic mantling processes, current state comparison to primitive carbonaceous chondrites, volatile composition link back to differential volatile loss in active comets |
| **Interstellar Objects**: *Planetesimal formed in an extra-solar disk.* | Volatile composition links to diversity in proto-planetary disks |
| **Near-Earth Asteroids (NEA)**: *Fragments of main-belt asteroids, transported via Yarkovsky Effect and resonances* | Representativeness of meteorite/sample return samples, biases of meteorite collection |
| **Differentiated Asteroids (ice/rock)**: *Planetesimal or planetary embryo, transported to asteroid belt from giant planet region. Ceres is archetype, is a relict Ocean World with surviving ice in upper layers of regolith.* | Early solar system heat sources, subsurface oceans, aqueous processes. |
| **Differentiated Asteroids (rock/metal)**: *Planetesimal or planetary embryos formed in inner solar system, having experienced igneous processes. Can include exposed/exhumed mantle/core fragments.* | Early solar system heat sources, first inner solar system solids |
| **Large asteroids with no collisional family**: *Planetesimal or planetary embryo, not contributing to meteorite collection and plausibly unrepresented.* | Diversity of material available for accretion to planets, compositional diversity of planetesimals |
| **Asteroid Family Members**: *Fragments of parent bodies of various properties* | Heterogeneity of parent bodies, meteorite/NEO delivery, studies for which compositional control is important |
| **Carbonaceous, Primitive**: *Material akin to carbonaceous chondrites, formed in giant planet region, experienced aqueous alteration/metamorphism but not differentiation.* | Nature of non-ice material in outer solar system, delivery of prebiotic material to terrestrial planets. |
| **Non-carbonaceous, Primitive**: *Material akin to ordinary chondrites, formed in terrestrial planet region, potentially experienced thermal metamorphism but not differentiation.* | Bulk compositions of terrestrial planets, processes on other anhydrous, silicate, airless surfaces (lunar, mercurian) |
| **Trojan/Hilda Asteroids**: *Transported from TNO region during planetary migration, captured into Jovian resonances.* | Large-scale planetesimal transport, non-volatile composition of TNOs. |

surprising that when small bodies are even coarsely divided into broad groupings as has been done in Table 1, that so many distinct science objectives arise. Important science questions emerge within each of these broad groups as well, such as when comparing quiescent stable Centaurs to active outbursting ones, or observing the significant surface and bulk differences between Bennu and Ryugu, despite similar sizes and compositions. This is closely similar to questions emerging within the broad groups of "Icy Satellites" when comparing the superficially similar Europa and Enceladus, or when contrasting the Martian poles with one another.

**The variety found in the small bodies population**
An impressively large range in sizes is not sufficient to warrant interest or investigation, but the small bodies also cover an extremely large range of formation locations and conditions. Meteorite samples, which come from asteroidal parent bodies in all but unusual cases, suggest ~150 distinct parent bodies exist in today's asteroid belt. Cousins of the planetesimals that accreted to ultimately form the inner planets still can be found in the asteroid belt. Planetesimals that formed among the giant planets are also thought to be found in the asteroid belt, transported during a period of planetary migration. The region beyond Neptune is a third location of small body formation, currently containing objects that formed either in-situ or inside of Neptune's current location. From this reservoir some objects have been transported to orbits throughout the outer planets (Centaurs) and as far inwards as the outer asteroid belt or into the population of Jupiter Family Comets (JFC). Even the most distant small body reservoir in the solar system reservoir, the Oort cloud, is populated by objects that formed throughout the current giant planet region. It is thought that many outer planet satellites have been captured from small body populations, and in some cases have been disrupted while in orbit to form satellite families (for instance, the Himalia or Ananke families orbiting Jupiter). The moons of Mars, Phobos and Deimos, may have been similarly captured, or may represent surviving ejecta from a giant impact onto Mars—an origin that has also been proposed for the small population of olivine-rich objects orbiting in the main Asteroid belt and near the leading and trailing Lagrange points of Mars. Indeed, even objects formed outside of our Solar System are studied by small bodies experts, not only 1I/`Oumuamua and 2I/Borisov, but presolar grains found in meteorites and interstellar dust moving through the Solar System.

The forces experienced by small bodies are far more varied than those that are found on large ones, and any conception that these objects are identically-inert lumps of rock and/or ice is as mistaken as conceiving of Jupiter, the Sun, and Betelgeuse as similar spheres mostly made of hydrogen (Figure 1). While endogenic processes are not as important on small bodies as they are on large ones, they are not absent. Tenuous atmospheres and exospheres are found around some objects, some present seasonally or episodically. Ring systems have been found around some outer solar system objects (TNO Haumea, Centaurs Chariklo and possibly Chiron) via occultation measurements, and their prevalence in the population at large is basically unknown. Cryovolcanic structures have been found on Ceres, and large-scale igneous processes have occurred on Vesta and on several meteorite parent bodies. Indeed, the first and only planetary cores that humanity will visit in the foreseeable future (if not ever) will be the metallic cores of small bodies, evidence of planetary differentiation and exposed by massive collisions.

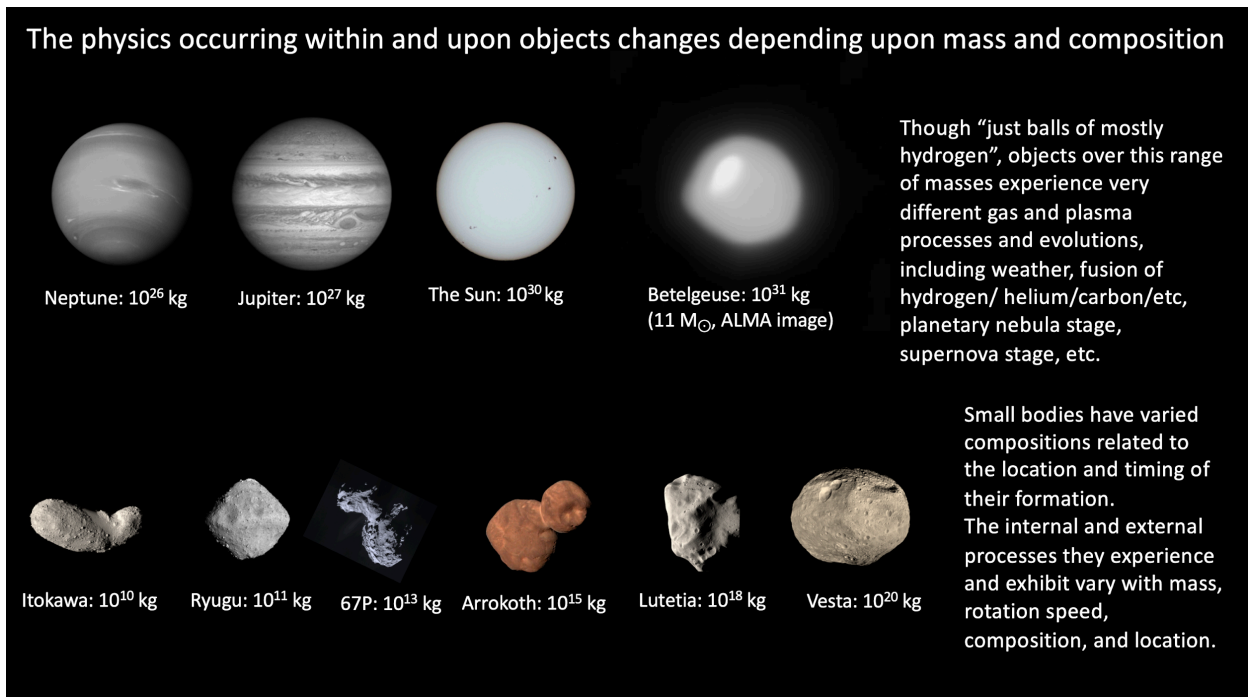

**Figure 1.** *While stars and giant planets could all be characterized as "hydrogen-dominated masses in hydrostatic equilibrium", the processes they experience differ greatly based on their masses, and investigation of Neptune is not justification for considering study of M-type giants to be underway. Similarly, study of a member of one subgroup of the small bodies, which cover a much wider range of masses (as well as other important physical properties), is not justification for failing to study other subgroups, nor for investigating the trends within these subgroups.*

In addition to endogenic processes, small bodies experience an array of exotic exogenic forces too small to be detected on planets or large satellites but able to change the surfaces, spins, shapes, and even orbits of small bodies. Heat not only drives activity in comets via sublimation of volatiles, but also drives such activity in a population of icy asteroids. Thermal forces also directly drive activity via dehydration of hydrated minerals in low-perihelion objects like 3200 Phaethon and indirectly through the "YORP Torque", which can increase the spins of km-scale and sub-km objects until they disrupt via rotational fission. Numerous comets have been observed to fragment or split. While the cause of splitting and implications of fragment compositions may not be fully understood, it is ultimately tied to heating, either solar or tidal.

**Connections between small bodies and other Solar System objects and populations**

As might be expected, studies of small body populations and their constituent objects also have deep connections with and implications for larger objects throughout the solar system and beyond. The distribution of small bodies in their orbits is the most powerful tracer of planetary migration in the early history of our Solar System, and meteorite samples provide direct evidence of planetesimal compositions and formation conditions for the terrestrial planets that are unavailable from any other set of objects.

The variety of airless body compositions and solar distances provide tests and additional constraints for theories about lunar processes such as space weathering, electrostatic levitation of regolith, and solar wind implantation. The contribution of infall from small bodies is important to many aspects of lunar science: without knowledge of carbonaceous chondrite compositions, interpretations of Apollo lunar regolith compositions would have been much less secure, and the ice in permanently shadowed regions on the Moon and Mercury is thought to have been delivered by one or more cometary impacts.

Small bodies studies are relevant to Mars science, as well. Questions about the formation of Mars directly led to the development of several dynamical scenarios, including the Grand Tack, which are most directly tested by studying the distribution of asteroid orbits and compositions. There is evidence from dynamical and collisional studies that the populations of small bodies associated with Mars, Phobos and Deimos and the Mars Trojan asteroids, may be Mars ejecta—the former pair forming from a circum-Mars disk after a giant impact, and the latter group debris after a basin-forming impact that also put olivine-rich objects elsewhere in the inner asteroid belt.

Icy worlds in the outer solar system also have close connections to the small bodies, and interpretations of their natures are reliant on small bodies research. Some of the so-called "Ocean Worlds" are members of small bodies populations, and serve as natural, "free-range" comparisons to similar-sized objects in orbit around giant planets, providing constraints for understanding how much of the character of these bodies is shaped by the systems where they reside. Similarly, while we lack meteorites that we are confident come from parent bodies currently in the outer solar system, we expect the Ocean Worlds to be constructed of planetesimals that are the outer-solar-system counterparts to the asteroids. Indeed, given current paradigms about the transport of material during planetary migration, it seems likely that the non-ice material in low-albedo asteroids in the main belt and Trojan regions is largely the same as what was found in outer-solar-system building blocks. While more volatile ices that were accreted by proto-Ocean Worlds are absent from present-day asteroids due to sublimation, we can expect them to still be present in cometary, Centaur, and TNO populations.

Due to their dynamical history, the small icy planetesimals afford us the unique opportunity to characterize how their surfaces and compositions have evolved since formation. The difference between the seemingly smooth surfaces of the TNO Arrakoth and the deep pits or sharp scarps on 67P/Churyumov-Gerasimenko could not be more stark. Only by fully exploring the diversity of TNOs, JFCs and the Centaurs in between will we understand how one transforms to the other as orbital billiards with the giant planets sends the small icy bodies throughout the solar system transforming them via outbursts of gas, dust and sometimes persistent coma production along the way. Inside ~4 AU this activity is understood to be driven primarily by water sublimation, but which of our hypotheses explain the activity and thus surface and potentially chemical evolution of Centaurs and LPCs when they are beyond the water regime? This is essential context for drawing the line from comet composition and structure back to disk composition and planetesimal formation.

The small bodies are central players in astrobiology. They were certainly key carriers of prebiotic material to the terrestrial planets, though the relative proportions brought by cometary objects vs. by carbonaceous asteroids is still a matter of debate. There may also have been habitable environments in small bodies, particularly those that are also Ocean Worlds. We can anticipate astrobiologically-driven exploration of Ceres in this coming decade.

All of these ways in which small bodies studies are relevant to other objects in this Solar System also apply to exoplanetary studies. Planetesimals will only ever be observed, let alone explored in situ or sampled, in this Solar System. Indeed, the best opportunities for obtaining macroscopic samples of objects formed in exosystems at all, let alone anytime in this century, will be by visiting a small body discovered via near-Earth object surveys while traversing our Solar System, and what exoplanetary material we currently have is found in meteorites. More indirectly, insights into the formation and evolution of our Solar System gained from the study of small body orbits and compositions provide constraints for the formation and evolution of exosystems.

**The interdisciplinary nature of small bodies studies**

The diversity of the small bodies is matched by the diversity of approaches used to study them, mirroring the diversity of approaches found in planetary science as a whole. Laboratory geochemistry of meteorite samples, photogeology of mission targets, shape modeling using radar data, calculations of the evolution of organic matter in subsurface oceans, modeling of coma chemistry and dynamics, ground-based telescopic imaging, emission spectroscopy, and reflectance spectroscopy, dynamical calculations of orbit evolution over billions of years, among many other things, all fall under the aegis of small bodies science.

In the spirit of this interdisciplinary nature, we recommend that the decadal survey consider the critical role of team dynamics, equity, diversity, inclusion, and accessibility in planetary science. As demonstrated in this report, the study of small bodies requires drawing on perspectives spanning the gamut of planetary science, geoscience, astronomy, technology, engineering, and beyond. Studies of scientific teams have repeatedly demonstrated the importance of an integrated approach, where team members with diverse expertise develop synergies between their specialties and resources that result in an end product that adds up to more than the sum of its parts.

Additionally, it is critical that the planetary science community fosters an interdisciplinary, diverse, equitable, inclusive, and accessible environment. We, along with many other white paper authors, strongly encourage the decadal survey to consider the state of the profession and the issues of equity, diversity, inclusion, and accessibility—not as separable issues, but as critical steps on the pathway to understanding the small bodies population and the entire solar system.

**Recommendation**

The great diversity of the small bodies population and their centrality in answering such a broad range of fundamental planetary science and astrobiology questions about the origin and evolution of our Solar System and exoplanetary systems warrants continued in-depth exploration and research. We urge the Decadal Survey to emphasize the diversity of small bodies and recommend that funding agencies commit to exploring all of them on their own merits, and to recognize that just as simultaneous investments in missions to different outer planet satellites or in simultaneous missions to Mars' orbit and surface strengthens the science return from each of them, simultaneous investments in missions across the range of small body populations strengthens their science return, and does so for related fields outside of small bodies science.

We advocate for a deep and ongoing program of missions, telescopic observations, R&A funding, and student support that respects this diversity.